\documentclass[showpacs,superscriptaddress,floatfix,twocolumn,prl]{revtex4}

\usepackage{amsmath}
\usepackage[dvips]{graphicx}
\usepackage{epsfig}
\usepackage{tabularx}

\begin{document}

\title{Classical signature of ponderomotive squeezing in a suspended mirror resonator}

\author{Francesco~Marino}
\affiliation{Dipartimento di Fisica, Universit\`a di Firenze, INFN
Sezione di Firenze, and LENS,\\ Via Sansone 1, I-50019 Sesto Fiorentino
(FI), Italy}

\author{Francesco~S.~Cataliotti}
\affiliation{Dipartimento di Energetica, Universit\`a di Firenze, INFN
Sezione di Firenze, and LENS,\\ Via Santa Marta 3, I-50139 Firenze, Italy}

\author{Alessandro~Farsi}
\affiliation{Dipartimento di Fisica, Universit\`a di Firenze, INFN
Sezione di Firenze, and LENS,\\ Via Sansone 1, I-50019 Sesto Fiorentino
(FI), Italy}

\author{Mario~Siciliani~de~Cumis}
\affiliation{Istituto Nazionale di Ottica Applicata, and INFN
Sezione di Firenze, \\ L.go E. Fermi 6, I-50125 Firenze, Italy}

\author{Francesco~Marin}
\affiliation{Dipartimento di Fisica, Universit\`a di Firenze, INFN
Sezione di Firenze, and LENS,\\ Via Sansone 1, I-50019 Sesto Fiorentino
(FI), Italy}

\date{\today}
\begin{abstract}
The radiation pressure coupling between a low-mass moving mirror and an incident light field has been experimentally studied in a high-finesse Fabry-Perot cavity. Using classical intensity noise in order to mimic radiation pressure quantum fluctuations, the physics of ponderomotive squeezing comes into play as a result of the opto-mechanical correlations between the field quadratures. The same scheme can be used to probe ponderomotive squeezing at the quantum level, thus opening new routes in quantum optics and high sensitivity measurement experiments.
\end{abstract}


\maketitle

The observation of quantum phenomena in the mechanical interaction between light and macroscopic objects (quantum opto-mechanics) is a fundamental problem in quantum physics
and in experiments involving highly sensitive measurements. 
In particular, the possibility of exploiting the opto-mechanical coupling between light and mirrors to generate non-classical states of light has attracted most of the attention.
In suspended mirror resonators, the radiation pressure creates correlations between the field quadratures: amplitude fluctuations move one mirror through radiation pressure, which leads to phase fluctuations. In this way one can obtain radiation with fluctuations in one quadrature smaller than the Standard Quantum Limit (SQL) (ponderomotive squeezing) \cite{mancini94,sq1,sq3,sq4}. Experimental systems enabling the observation of this phenomenon would also allow other, even more relevant investigations, such as quantum non-demolition measurements of the field quadratures \cite{qn1,qn2,qn3}, and eventually the creation of entangled states of light and one or more oscillators~\cite{entanglement0,entanglement1,entanglement2,entanglement3} or the observation of the quantum ground state of a macroscopic mechanical oscillator \cite{qgs1,qgs2}.

Despite the extreme relevance of the mentioned proposals and theoretical studies, quantum opto-mechanical phenomena involving macroscopic systems have never been experimentally observed. This is mainly due to the intrinsic weakness of the radiation pressure quantum fluctuations with respect to classical noise sources: acoustic and thermal to name just a few. An impressive sequence of experimental advances has been recently reported, mainly focused on radiation pressure cooling \cite{Cohadon99,Kleckner06,Gigan06,Arcizet06,corbitt2,Thompson08,Schliesser08,aspel,Park09}. This technique allowed to approach the observation of a macroscopic oscillator in its ground state. Moreover, progress in the areas of nano-technology and of Micro-Electro-Mechanical Systems (MEMS)  approaches the realization of suitable experimental platforms.

In this work we have studied experimentally a high-finesse Fabry-Perot cavity with a MEMS end-mirror. Although the opto-mechanical properties of our device are not presently sufficient to allow the direct observation of quantum phenomena, the physics of ponderomotive squeezing comes into play by adding intensity noise to the intracavity field. Classical fluctuations in similar systems have been recently exploited to show classical analogs of quantum noise cancellation \cite{mow}, back-action reduction \cite{can} and opto-mechanical correlations between two optical beams \cite{ver}. In this context, our results serve the dual purpose to make experimentally evident the physics behind ponderomotive squeezing, and to prove the validity of our experimental scheme in view of a first experiment entering the quantum regime.
\begin{figure}
\begin{center}
\includegraphics*[width=1.0\columnwidth]{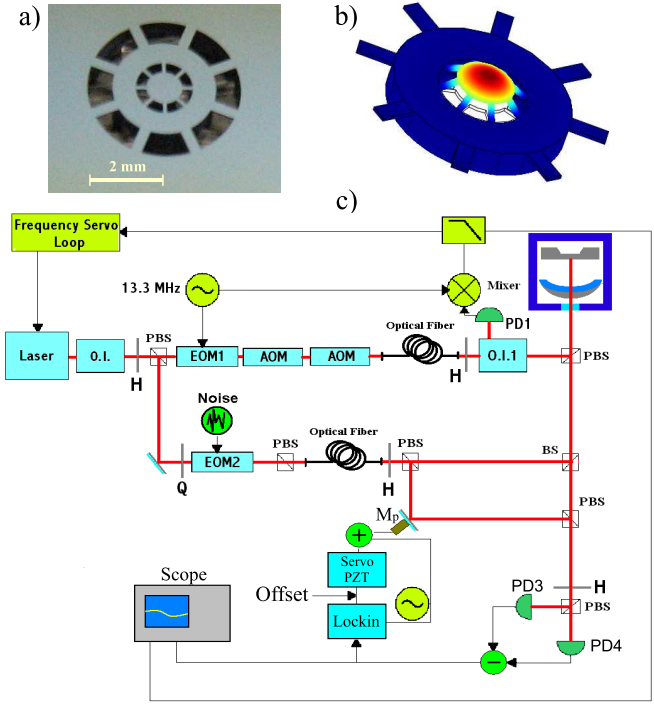}
\end{center}
\caption{a) Image of the double wheel oscillator. b) FEM calculation of the fundamental mode. c) Scheme of the experimental apparatus. O.I.: optical isolator; AOM: acousto-optic modulator; EOM: electro-optic modulator; H: half-wave plate; Q: quarter-wave plate; PD: photodiode; PBS: polarizing beam-splitter; BS: beam-splitter.}
\vspace{-.3cm}
\label{figu1}
\end{figure}

Our MEMS sample, realized on a 400~$\mu$m silicon substrate, consists of an array of double wheel oscillators like the one shown in Fig.~\ref{figu1}a. The central mass and the arms are 45~$\mu$m thick, while the intermediate ring mass is 400~$\mu$m thick. These particular structures has been chosen in order to reduce the mechanical coupling between the main oscillator and high-frequency modes of the whole sample. On the front side of the wafer, a deposition of alternate Ta$_2$O$_5$/SiO$_2$ quarter-wave layers for a total thickness of about 5~$\mu$m provides the highly reflective coating.
One of these double-oscillators is used as end mirror of a $12.2$~mm long Fabry-Perot cavity with a $50$~mm radius silica input mirror (transmissivity $\mathcal{T}$=110 ppm) operating in a vacuum chamber at $5\cdot10^{-4}$~Pa.
We obtained a cavity finesse of $\mathcal{F}$ = 10000, limited by the optical losses of the oscillating mirror. The size of the reflection dip is $38\%$ of the total power. A FEM (Finite Element Method) study predicts the frequency of the oscillator fundamental mode at $\sim 259$~kHz. The corresponding spatial deformation is displayed in Fig. \ref{figu1}b.

The experimental setup for the measurement of the phase and amplitude fluctuations of the intracavity field is sketched in Fig.~\ref{figu1}c. The light source is a cw Nd:YAG laser operating at $\lambda$=1064~nm. After a 40~dB optical isolator, the laser radiation is split into two beams. On the first one (locking beam), a resonant electro-optic modulator (EOM1) provides phase modulation at 13.3~MHz used for the Pound-Drever-Hall (PDH) detection scheme \cite{Drever}. The locking beam can be frequency shifted by means of two acousto-optic modulators (AOM) operating on opposite diffraction orders.
The intensity of the second beam (signal beam) is controlled by an electro-optic modulator (EOM2) followed by a polarizing beam splitter. Both beams are sent to the second part of the apparatus by means of single-mode, polarization maintaining optical fibers. 
The two beams are overlapped with orthogonal polarizations in a polarizing beam-splitter and sent to the optical cavity. The reflected locking beam, on its back path, is deviated by the input polarizer of a second optical isolator (O.I.1) and collected by a photodiode (PD1) for the PDH detection. This PDH signal is used for laser frequency locking, while the signal beam is used for the squeezing experiment. Since the cavity is birefringent (by $\approx$~100~kHz), the two beams are frequency-shifted by the same quantity so that they both match the cavity resonance. Such frequency mismatch has the advantage to prevent any spurious interference and reduce the cross-talk in the photo-detection.

The reflected signal beam, suitably attenuated, is mixed with a strong local oscillator field, derived from the same beam before sending it to the cavity, and monitored by homodyne detection. At this point, the ratio between the signal and the local oscillator intensities is 1/100. The homodyne detector consists of a half-wave plate and a polarizing beam-splitter dividing the beam into two equal parts sent to the photodiodes PD3 and PD4 whose outputs are then subtracted. The dc component of the resulting difference current is proportional to $\cos \phi$, where $\phi$ is the optical phase difference between the reflected signal and the local oscillator (homodyne angle). The piezoelectric-driven mirror $M_p$ is modulated by a sinusoid generated by the internal oscillator of a digital lock-in amplifier. The difference photocurrent is then demodulated, integrated and fed back onto $M_p$. This procedure fixes $\phi$ in such a way to detect the amplitude quadrature. By adding an offset $V_{off}$ to the error signal, the homodyne angle can be varied, to detect a selected field quadrature $\delta X_{\phi}$, i.e., a mixture of amplitude ($\delta X_1$) and phase ($\delta X_2$) fluctuations~\cite{Gerry}:
\begin{equation}
\delta X_{\phi} = \delta X_1 \cos\phi - \delta X_2 \sin\phi  \, .
\label{quadratura}
\end{equation}
As explained later, when the laser is locked to the cavity, the noise spectral density $S_{\infty}(V_{off})$ at frequencies far from the mechanical resonance
is proportional to $\cos^2\phi$. Therefore, the angle $\phi$ can be determined by measuring $S_{\infty}(V_{off})$, at different offset values. $\phi$ is then obtained by inverting the expression $S_{\infty}(V_{off})/S_{\infty}(0)=\cos^2\phi$.

\begin{figure}
\begin{center}
\includegraphics*[width=1.0\columnwidth]{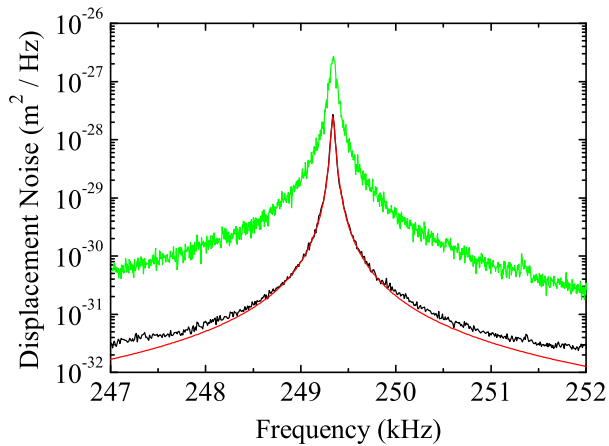}
\end{center}
\caption{Spectral density of the cavity length fluctuations measured without (black) or in presence of the signal beam with added intensity noise (green). The red curve is a thermal noise fit through the expression of $S^{th}_x$ given in the text, at room temperature.}
\vspace{-.3cm}
\label{figu2}
\end{figure}

A typical displacement noise spectrum, as obtained from the PDH signal after having shut down the signal beam and with the laser weakly locked on the cavity, is reported in Fig. \ref{figu2}.
The signal is calibrated by modulating the frequency of the laser through its internal piezoelectric crystal, with a depth smaller than the Fabry-Perot cavity linewidth, and using a phase-sensitive detection on the PDH signal. In the case of thermal noise excitation, the displacement spectrum at temperature $T$ is given by $S^{th}_x(\omega) = \frac{4kT}{\omega} \textrm{Im}~\chi(\omega)$ where
\begin{equation}
\chi(\omega) = \frac{1}{m (\omega_M^2-\omega^2-i \omega\omega_M / Q)} \; \\
\label{eq1}
\end{equation}
is the end-mirror susceptibility, $\omega_M / 2 \pi$ its resonance frequency and $m$ its effective mass.
By fitting the data through the expression of $S^{th}_x$ at room temperature we infer $\omega_M / 2 \pi \sim$~249300~Hz, $Q \sim$~5500 and $m \sim 1.1~10^{-7}$~Kg. A mass value of $m \sim 9.6~10^{-8}$~Kg, in excellent agreement with the previous one, is found by extracting the susceptibility from the response of the mirror displacement to  amplitude modulated radiation (the method is described in detail in Ref.~\cite{Decumis}). This allows us to confirm the thermal nature of the spectrum around the mechanical resonance.

In order to enhance its amplitude fluctuations, the signal beam is intensity modulated by EOM2 to produce classical intracavity radiation pressure noise. In this position (i.e., before the separation of the local oscillator) the phase noise added by EOM2 is not relevant for the experiment.
The electronic noise signal driving EOM2 is generated by an arbitrary waveform generator, 
whose output (initially with a flat spectrum up to $\sim 10$~MHz) is band-pass filtered
with a bandwidth of about 15 kHz centered approximately at the mechanical resonance. Therefore, it can be safely considered as white noise around the mechanical peak (width $\sim$ 45 Hz). The amplitude of the electronic noise and the laser power (at the cavity input, $P_{in} \sim$ 5.6 mW) are chosen so that the force noise due to fluctuations of the intracavity radiation pressure is 10 dB larger than thermal noise force $S^{th}_F=-\frac{4kT}{\omega} \textrm{Im}\frac{1}{\chi}$. As shown in Fig.~\ref{figu2}, the displacement noise spectrum measured in the PDH signal is indeed enhanced by 10~dB and, around the peak, the radiation pressure is now the dominant noise source. In this way, we essentially reproduce, at a higher noise level, the physical conditions in which ponderomotive squeezing takes place.

\begin{figure}
\begin{center}
\includegraphics*[width=1.0\columnwidth]{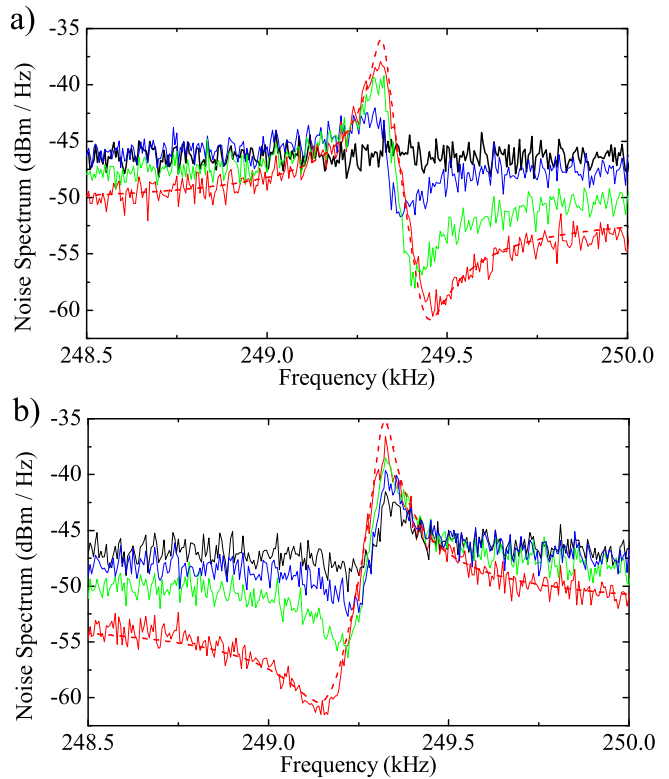}
\end{center}
\caption{Noise spectra of the reflected signal beam as the homodyne angle is varied. a) $\phi$ =0 (black trace), $\phi =-20^{\circ}$ (blue trace), $\phi =-44^{\circ}$ (green trace), $\phi =-56^{\circ}$ (red trace). b) $\phi =26^{\circ}$ (black trace), $\phi =35^{\circ}$ (blue trace), $\phi =45^{\circ}$ (green trace), $\phi =62^{\circ}$ (red trace). Dashed lines are used for the theoretical curves.}
\vspace{-.3cm}
\label{figu3}
\end{figure}
For zero offset we simply detect the amplitude noise of the signal beam (black trace in Fig. \ref{figu3}a) which, in our classical analogy, plays the role of the shot noise. For nonzero offset values, the noise spectrum consists of a mixture of two contributions: amplitude fluctuations, which also act on the mirror via radiation pressure, and phase fluctuations induced by the consequent mirror motion. Far from the mechanical resonance, only the first contribution is relevant, and the spectral density is simply proportional to $\cos^2 \phi$, keeping a constant signal-to-noise ratio as expected for classical noise. Approaching the frequency region where the opto-mechanical effect becomes important, the resulting noise spectrum falls below the "shot-noise" level on one side of the mechanical resonance, while increases on the other side. This can be intuitively understood by considering that before resonance, force (i.e., field amplitude) fluctuations and displacement (i.e., reflected field phase) fluctuations are in phase. Above resonance, when the real part of the susceptibility has opposite sign, they are in anti-phase. The detected quadrature (see Eq.~(\ref{quadratura})) reflects this behavior. Indeed, by changing the sign of the offset (i.e., of the homodyne angle) the regions of positive and negative interference are inverted (see Fig. \ref{figu3}b). We observe a maximum noise reduction of 9~dB thanks to the ponderomotive effect, limited by the mirror thermal noise. A calculation of the expected signals was performed following Ref.~\cite{sq1} and adding the homodyne detection, the cavity losses and the laser extra-noise. Two examples of the theoretical curves, obtained with the previously measured opto-mechanical parameters, are shown in Fig. \ref{figu3}. The agreement with the experiment is excellent.

\begin{figure}
\begin{center}
\includegraphics*[width=1.0\columnwidth]{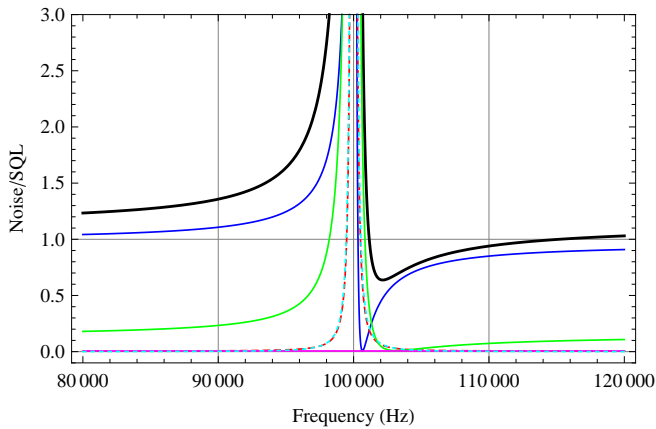}
\end{center}
\caption{Noise spectral density, normalized to its Standard Quantum Level, calculated for the parameters given in the text and a homodyne angle of $10^{-4}$ rad (thick black line). Also shown are the different contributions to the total noise: vacuum fluctuations entering through cavity losses (blue line) and in the beam-splitter that divides signal and local oscillator (dashed red line); laser amplitude noise (green); laser frequency noise (violet); mirror thermal noise (light blue, dashed).}
\vspace{-.3cm}
\label{figteo}
\end{figure}
The measurements to show ponderomotive squeezing in the quantum regime would be similar to those here performed to characterize its classical counterpart. In order to produce field fluctuations below the quantum noise level, the essential requirement is that the radiation pressure force noise due to the quantum fluctuations of the input field (amplified in the cavity) must dominate over thermal force noise. In terms of the involved physical quantities, $\,\,\hbar \omega_L P_{in}\frac{4}{c^2} \mathcal{T}^2 \left(\frac{\mathcal{F}}{\pi}\right)^4 \gtrsim 2 k T \frac{m \omega_M}{Q}$ ($\omega_L/2\pi$ is the laser frequency).  On the other hand, the input field do not need to be strictly shot-noise limited, since excess amplitude fluctuations are cancelled, on a suitably chosen quadrature, by the same mechanism that produces quantum noise squeezing. Also laser frequency noise and wideband mirror displacement noise are not very critical parameters since, close enough to the resonance peak, the effect of force noise on the oscillator is strongly enhanced. The wideband frequency and displacement noise only limits the available range of homodyne phase and spectral region where the squeezing can be observed. As an example, we report in Fig.~\ref{figteo} the expected noise, normalized to its quantum level, that can be observed in an experiment with realistic parameters. The experiment should be performed at cryogenic temperature~\cite{aspel,Decumis}, and we consider cavity optical losses of 40~ppm (that we could indeed achieve with an accurate cleaning procedure), $\mathcal{T}=50$~ppm, a cavity length of 6~mm (with a free-spectral-range still covered by the laser tuning range), $P_{in}=30$~mW, and slightly adjusted parameters of the mechanical oscillator, with $m=5\cdot 10^{-8}$~kg and $\omega_M/2\pi=100$~kHz. A mechanical quality factor of $Q\sim 10^5$ is expected considering the loss angles of the coating and the silicon structure, weighted by their respective thickness~\cite{Geppo}. In the figure we also report the different contributions to the total noise, that include laser amplitude fluctuations higher than the shot-noise level by a factor of 5 \cite{conti}, and laser frequency noise of 1~Hz$^2$/Hz, equivalent to a displacement noise of $10^{-33}$~m$^2$/Hz. As shown, an observable noise reduction of about 2~dB is within the possibilities of our apparatus with reasonable improvements. We must remark that such an high intracavity power is very challenging from the point of view of the dynamic stability \cite{Arcizet06,sq1,Marino}, however some degree of squeezing is expected with the above parameters even for an input power reduced down to 2~mW. A further critical point is the choice and stability of the homodyne angle, that must be at the level of 10$^{-4}$~rad. This requires: i) a detection of the interference fringes (between the rather strong signal and local oscillator fields) with a signal-to-noise ratio of 10$^{-4}$, in a band extending to some tens kHz; ii) an active locking, on the same band, with a gain sufficient to reduce the in-loop fluctuations at the same level, corresponding to a 0.1~nm stability of the length of the interferometer arms. Both requirements are reasonably obtainable in a refined apparatus, according also to our preliminary tests.

In conclusion, we have studied a high-finesse Fabry-Perot cavity with a MEMS end-mirror
in which classical intensity noise has been added to the input field in order to enhance radiation pressure fluctuations. Whenever the displacement noise induced by radiation pressure becomes the dominant noise source, the physics of ponderomotive squeezing takes place as a result of the (classical) opto-mechanical correlations between the field quadratures. Similarly in the quantum regime, the same scheme is expected to allow the generation of squeezed light. This effect would provide the first demonstration of a quantum opto-mechanical effect in a macroscopic system, opening the way to an extremely promising experimental investigation of a field that involves the foundations of quantum mechanics.


\begin{thebibliography}{99}
\bibitem{mancini94} S. Mancini, P. Tombesi, Phys. Rev. A {\bf 49}, 4055 (1994).
\bibitem{sq1} C. Fabre {\it et al.}, Phys. Rev. A {\bf 49}, 1337 (1994).
\bibitem{sq3} S. Bose, K. Jacobs, P. L. Knight, Phys. Rev. A {\bf 56}, 4175 (1997).
\bibitem{sq4} T. Corbitt {\it et al.}, Phys. Rev. A {\bf 73}, 023801 (2006).
\bibitem{qn1} K. Jacobs, P. Tombesi, M. J. Collett, D. F. Walls, Phys. Rev. A {\bf 49}, 1961 (1994).
\bibitem{qn2} M. Pinard, C. Fabre, A. Heidmann, Phys. Rev. A {\bf 51}, 2443 (1995).
\bibitem{qn3} A. Heidmann, Y. Hadjar, M. Pinard, Appl. Phys. B {\bf 64}, 173 (1997).
\bibitem{entanglement0} S. Mancini, V. Giovannetti, D. Vitali, P. Tombesi, Phys. Rev. Lett. {\bf 88}, 120401 (2002).
\bibitem{entanglement1} W. Marshall, C. Simon, R. Penrose, D. Bouwmeester, Phys. Rev. Lett. {\bf 91}, 130401 (2003).
\bibitem{entanglement2} M. Pinard {\it et al.}, Europhys. Lett. {\bf 72}, 747 (2005).
\bibitem{entanglement3} D. Vitali {\it et al.}, Phys. Rev. Lett. {\bf 98}, 030405 (2007).
\bibitem{qgs1} R.G. Knobel and A.N. Cleland, Nature {\bf 424}, 291 (2003). 
\bibitem{qgs2} M.D. LaHaye, O. Buu, B. Camarota, and K.C. Schwab, Science {\bf 304}, 74 (2004).
\bibitem{Cohadon99} P.F. Cohadon, A. Heidmann, and M. Pinard, Phys. Rev. Lett. {\bf 83}, 3174 (1999).
\bibitem{Kleckner06} D. Kleckner and D. Bouwmeester, Nature {\bf 444}, 75 (2006).
\bibitem{Gigan06} S. Gigan {\it et al.}, Nature {\bf 444}, 67 (2006).
\bibitem{Arcizet06} O. Arcizet {\it et al.}, Nature {\bf 444}, 71 (2006).
\bibitem{corbitt2} T. Corbitt {\it et al.}, Phys. Rev. Lett.  {\bf 99}, 160801, (2007).
\bibitem{Thompson08} J.D. Thompson {\it et al.}, Nature {\bf 452}, 72 (2008).
\bibitem{Schliesser08} A. Schliesser {\it et al.}, Nat. Phys. {\bf 4}, 415 (2008); Nat. Phys. {\bf 5}, 509 (2009).
\bibitem{aspel} S. Gr\"oblacher {\it et al.}, Nat. Phys. {\bf 5}, 485 (2009).
\bibitem{Park09} Y-S. Park and H. Wang, Nat. Phys. {\bf 5}, 489 (2009).
\bibitem{mow} C.M. Mow-Lowry {\it et al.}, Phys. Rev. Lett. {\bf 92}, 161102 (2004).
\bibitem{can} T. Caniard {\it et al.}, Phys. Rev. Lett. {\bf 99}, 110801 (2007).
\bibitem{ver} P. Verlot {\it et al.}, Phys. Rev. Lett. {\bf 102}, 103601 (2009).
\bibitem{Drever} R.W.P.~Drever {\it et al.}, Appl. Phys. B {\bf 31}, 97 (1983).
\bibitem{Gerry} C.C. Gerry and P.L. Knight, {\it Introductory Quantum Optics}, Cambridge University Press, Cambridge, 2005.
\bibitem{Decumis} M Siciliani de Cumis {\it et al.}, J. Appl. Phys {\bf 106}, 013108 (2009).
\bibitem{Geppo} D. R. M. Crooks {\it et al.}, Class. Quantum Grav. {\bf 19}, 883 (2002); S. D. Penn {\it et al.}, Class. Quantum Grav. {\bf 20}, 2917 (2003).
\bibitem{conti} L. Conti, M. De Rosa, F. Marin, Appl. Opt. {\bf 39}, 5732 (2000).
\bibitem{Marino} F. Marino, F. Marin, Phys. Lett. A {\bf 364}, 441 (2007).
\end{thebibliography}
\end{document}